\RequirePackage{fix-cm}
\documentclass[smallcondensed]{svjour3}     % onecolumn (ditto)
\smartqed  % flush right qed marks, e.g. at end of proof
\usepackage{graphicx}
\begin{document}

\title{New resonant cavity-enhanced absorber structures for mid-infrared detector applications%\thanks{Grants or other notes
}

\titlerunning{New RCE absorber structures}        % if too long for running head

\author{Moshe Zohar \and Mark Auslender \and Lorenzo Faraone \and Shlomo Hava}


%\affiliation{
%\supscr{1}Department of Electrical and Computer Engineering, \\
%Ben Gurion University of the Negev POB 653, Beer Sheva 84105,
%Israel, \linkable{marka@ee.bgu.ac.il}, \\
%\supscr{2}Sami Shamoon College of Engineering, POB
%950, Beer Sheva
%84100, Israel \linkable{zoharmo@ee.bgu.ac.il} \\
%\supscr{3}School of Electrical, Electronic and Computer
%Engineering, The University of Western Australia, M018 35 Stirling
%Highway Crawley WA 6009, Australia}
\institute{M. Zohar \and M. Auslender \and S. Hava \at
              Department of Electrical and Computer Engineering, Ben Gurion University of the Negev, P.O. Box 653, 84105 Beer Sheva   \\
              Tel.:+972-8-6461583\\
              Fax: +972-8-6472949\\
              \email{zoharmo@ee.bgu.ac.il}           %  \\
           \and
           L. Faraone \at
              School of Electrical, Electronic and Computer Engineering, The University of Western Australia, M018 35 Stirling Highway Crawley WA 6009, Australia}
\date{Received: date / Accepted: date}

\maketitle

\begin{abstract}
A new dielectric Fabry-Perot cavity was designed for a resonant enhancing optical absorption by a thin absorber layer embedded into the cavity. In this cavity, the front mirror is a subwavelength grating with $\sim 100$\% retroreflection. For a HgCdTe absorber in a matching cavity of the new type, the design is shown to meet the combined challenges of increasing the absorbing efficiency of the entire device up to $\sim 100$ \% and reducing its size and overall complexity, compared to a conventional resonant cavity enhanced HgCdTe absorber, while maintaining a fairly good tolerance against the grating's fabrication errors.
\end{abstract}
\section{Introduction}
\label{intro}
It is well known, that incorporating a photosensitive or an optically active layer into a Fabry-Perot (FP) cavity enhances the efficiency of detection by or emission from the layer, as a consequence of the multiple reflections that occur between the cavity's mirrors. The enhancement is maximized if the round-trip phase, i.e. the phase difference between each succeeding reflection, satisfies the resonance condition
\begin{equation}
\delta_0\equiv \delta\left(\lambda_0\right)  = \frac{4\pi n_{\rm c}\left(\lambda_0\right)t_{c} }{\lambda_0} + \varphi_{\rm f}\left(\lambda_0\right) + \varphi_{\rm b}\left(\lambda_0\right) = 2\pi n. \label{resonant condition}
\end{equation}
Here $\lambda_0$ is a resonance wavelength $n_{\rm c}$ and $t_{\rm c}$ is the refractive index (RI) and length of the FP cavity, respectively, $\varphi_{\rm f}$ and $\varphi_{\rm b}$ are the reflection phase of the mirror that is further away from and adjacent to the illuminated side (hereafter referred to as the front and back mirror, respectively), and $n$ is an integer. Eq.(\ref{resonant condition}) is thumb rule accurate regardless of the type of flat-surface mirror being employed, whether it consist of thin metallic films, or is e.g. a distributed Bragg reflector (DBR) that is a stack of quarter-wave pairs of high/low (HL) RI dielectric layers \cite{heavens,knittl}.

Optical communication, interconnection and information processing systems require high-efficiency photodetectors (PDs). In response to this demand, the field of resonant-cavity-enhanced (RCE) PDs has steadily maturated over past two decades \cite{{unlu},deen}. For thermal PDs, the optical absorbance $A$ is an accurate measure of the efficiency, while for photodiodes and photo-conductors the key figure is the quantum efficiency $\eta$. For initial design of the RCE PDs, the adoption of $\eta \approx A$ has a widespread use \cite{{unlu},deen} since thus the analysis is framed into optics alone rather than being restricted to a specific PD technology. It appears \cite{unlu} that for obtaining $\eta \approx 100 \%$ it is at least necessary to have $\sim 100$\% resonant reflectivity from the front mirror, which is problematic for a DBR of any practical thickness because of the tight material requirements for epitaxial growth. Thus, it is relevant to consider an alternative replacement for the DBR mirror by a thin dielectric micro-optical component which can be $\sim 100$ \% reflective in a broad spectral band. A device that can exhibit such a reflection anomaly with a proper design \cite{havaus,{chang-a}} is a dielectric subwavelength grating structure.

This paper presents a theoretical study to incorporate a subwavelength grating as the front mirror in the FP cavity, for PD applications in the mid-wave infrared (MWIR) range. For a vertical-cavity surface-emitting laser, using an air-bridge subwavelength grating as a suspended top mirror, has been proposed before \cite{bissaillon,{kim}}. Only limited work on RCE PDs for operation in the MWIR range has been reported thus far. Currently this is an area of great interest due to the potential for obtaining the uncooled PDs.
%%---------------------------------------------------
\vspace{-0.05cm}
%%---------------------------------------------------
\section{Design considerations and simulation tools}
\label{sect:stat}
We adopted $\eta = A$ [see Sec. \ref{intro}], so RCE PDs with an absorber layer characterized by a complex RI, $n_{\rm a}+ik_{\rm a}$, were analyzed only optically. To compute the reflectance, transmittance and $A$ spectra of RCE PDs including a grating, we used the rigorous coupled wave analysis recast with an in-layer S-matrix propagation algorithm \cite{aushava1}. For DBR based PDs, a limiting case at zero grating groove width, which coincides with the impedance form \cite{knittl} of transfer matrix method, was employed. The optical spectra were computed in the range $4.215 \leq \lambda \leq 4.615 \;\mu{\rm m}$.

We put emphasis on the two issues: (i) optimizing the peak $A$ at $\lambda_0 = 4.415\; \mu$m and $\lambda_0 = 4.500\; \mu$m, which are kept fixed, while fulfilling Eq.(\ref{resonant condition}); (ii) seeking the designs with a high tolerance of the efficiency to errors in the grating fabrication. Interim design parameters serving as trial ones for computer aided optimization, will be obtained using the reflection phases [see Eq.(\ref{resonant condition})] of stand-alone mirrors \cite{unlu}, which means the mirrors put on an {\it infinite} medium of the same material as the cavity's material (CdTe in our designs) and irradiated {\it from} the substrate.

%%-------------------------------------------------------------------
\section{Conventional RCE PD structures}
\label{sect:RCE-conv}
The reported RCE HgCdTe-PDs \cite{{faraone1},faraone2,{faraone3}} have a standard configuration in which the CdTe FP cavity embeds a thin Hg$_{0.71}$Cd$_{0.29}$Te layer between two DBR mirrors [see in Fig.1(a)].
The HgCdTe/CdTe bilayers provide good lattice matching but the small difference in RIs of Hg$_{0.56}$Cd$_{0.44}$Te and CdTe requires $m = 30$, i.e. $\sim20\;\mu$m thick DBR mirror, in order to achieve the desired reflectivity \cite{faraone3}. Apart from enlarging the cavity, the deposition of such a thick multilayer inevitably produces growth defects which deteriorate the reflectance. In practice, such DBRs have been fabricated with $m=10$ \cite{kanev} and $m=7$ \cite{faraone3} and optimized with $m = 8$ for high temperature-operation RCE PDs \cite{sioma}.
%Figure 1 - RCE PDs structures (a) & (b)
\begin{figure}[h]
    \begin{center}
      \includegraphics[scale=0.22]{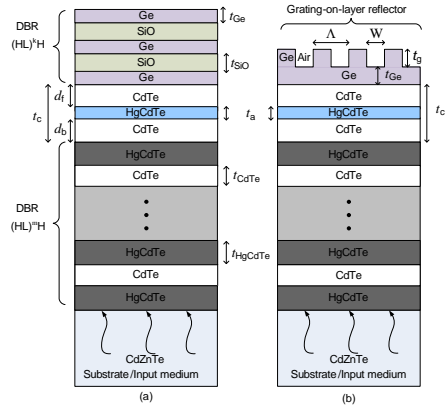}%0.25
      \caption{RCE HgCdTe-absorber structures with the: (a) (HL)$^k$H DBR and (b) grating front mirror; the back mirror in both is a (HL)$^m$H DBR; the irradiation is from a CdZnTe substrate.}
      \label{RCE PDs}
    \end{center}
\end{figure}
\vspace{-5pt}
We adopted a mid value of $m=15$, while due to \cite{{faraone1},{faraone2},{faraone3}} $k=2$, $t_{\rm a}=75\; {\rm nm}$. For a standard placement of the absorbing layer in the middle of the cavity ($d_{\rm f} = d_{\rm b}$), combining semianalytical calculation and manual optimization, we found the optimal cavity length for which $A=81.9\%$ at $\lambda_0=4.415\; \mu \rm m$ [RCE-S structure in Table \ref{tab:t2}]. The automatic optimization results in the same $t_{\rm c}$ as obtained manually. Further optimization, allowing for $d_{\rm f} \neq d_{\rm b}$, yields RCE-O$_a$ and RCE-O$_b$ structures with $\lambda_0=4.415$ and $\lambda_0 = 4.500\; \mu \rm m$, respectively [see in Fig.\ref{RCE conv.} and Table \ref{tab:t2}]. Figure \ref{RCE conv.} shows the $A$ spectra of the designed structures along with the reflectance ($R$) from the stand-alone front DBR mirror.

\vspace{-08pt}
\begin{table}[h]
\caption{The refractive indexes $N=n+ik$ used in the simulations}
\label{tab:t1}
\begin{center}
\begin{tabular}{|c|c|c|c|c|c|c|c|c|c|}
%% |l|l| to left justify each column entry
%% |c|c| to center each column entry
%% use of \rule[]{}{} below opens up each row
\hline
\rule[-1ex]{0pt}{3.5ex}  Wavelengths        & Ge            & SiO           & CdTe          & Hg$_{0.71}$Cd$_{0.29}$Te  & Hg$_{0.56}$Cd$_{0.44}$Te  & CdZeTe    \\%16.11.2011
\hline
\rule[-1ex]{0pt}{3.5ex} $4.415\; \mu \rm m$ & $3.9332$        & $1.78$          & $2.6695$        &$3.4826+ i 0.1477$     & $2.9709$                    & 2.6896    \\%16.11.2011
\hline
\rule[-1ex]{0pt}{3.5ex} $4.500\; \mu \rm m$ & $3.9325$        & $1.78$          & $2.6691$        &$3.4665+i\; 0.1425$     & $2.9697$                    & 2.6895    \\%new 16.11.2011
\hline
\end{tabular}
\end{center}
\end{table}

%%---------------------------------------------------
%\vspace{-0.01cm}
\vspace{-36pt}
%%---------------------------------------------------
\section{Grating mirror based RCE PD structures}
\label{sect:RCE-mod}
Figure 1(b) presents a modified RCE HgCdTe-PD, in which the front mirror is replaced by a one-dimensional (1-D) dielectric grating with period $\Lambda$, groove width $W$ and depth $t_{\rm g}$, while the absorber, back mirror and irradiation scheme remain the same as those of the structure in Sec. \ref{sect:RCE-conv}. The grating is designed as the stand-alone $\sim 100$ \% reflectivity mirror. Our design of the RCE absorber as a whole is done for the backside irradiation from CdZnTe substrate and is {\em not} suitable for the front irradiation from air [see Fig.\ref{RCE PDs}], because the grating mirror breaks down the irradiation symmetry inherent to the DBR based FP cavities.

%%---------------------------------------------------
\subsection{Polarization-selective RCE HgCdTe-absorber structures}
\label{subsect:RCE-A}
Earlier work \cite{havaus} and a recent paper \cite{chang-a} on designing grating mirrors have considered optical incidence from air, but here the stand-alone grating mirror is irradiated from CdTe. To get rid of the grating reflection orders, we imposed over all the computed wavelength range the subwavelength grating restriction $n_{\rm CdTe}\Lambda < \lambda$, which is tighter than for the incidence from air. Because of polarization sensitivity of the 1-D gratings, we first maximized the $R$ for the incidence from CdTe half-space, separately for the TE and TM polarization.
%%----------------------------------------------------------
\vspace{-12pt}
\begin{center}
\begin{figure}[h]
  \hfill
  \begin{minipage}[t]{.48\textwidth}
    \begin{center}
      \includegraphics[scale=0.38]{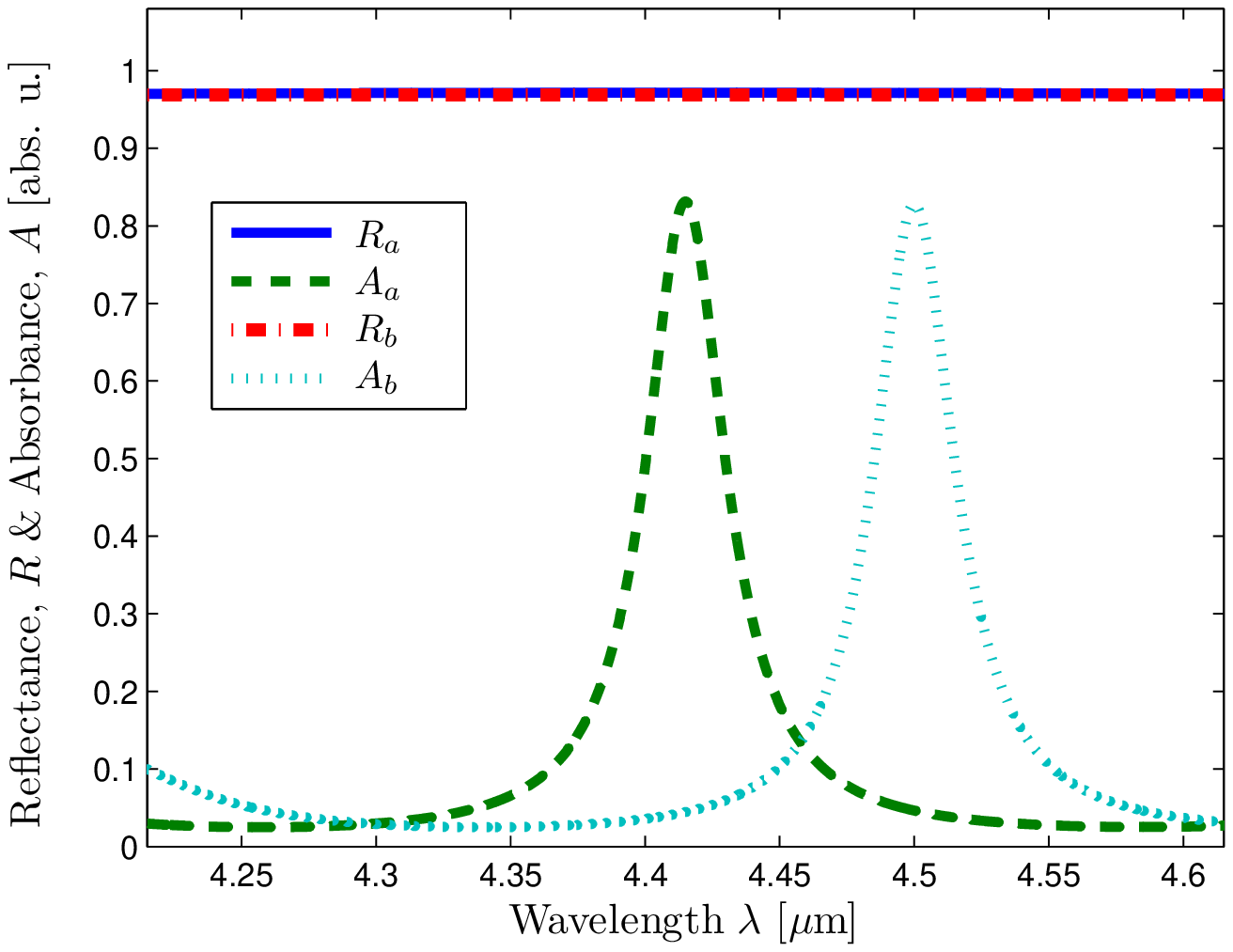}%0.38
      \caption{The spectra of reflectance from the front DBR mirrors and of absorbance for the RCE-O$_a$ and RCE-O$_b$ structures.}
      \label{RCE conv.}
    \end{center}
  \end{minipage}
  \hfill
  \begin{minipage}[t]{.48\textwidth}
    \begin{center}
      \includegraphics[scale=0.38]{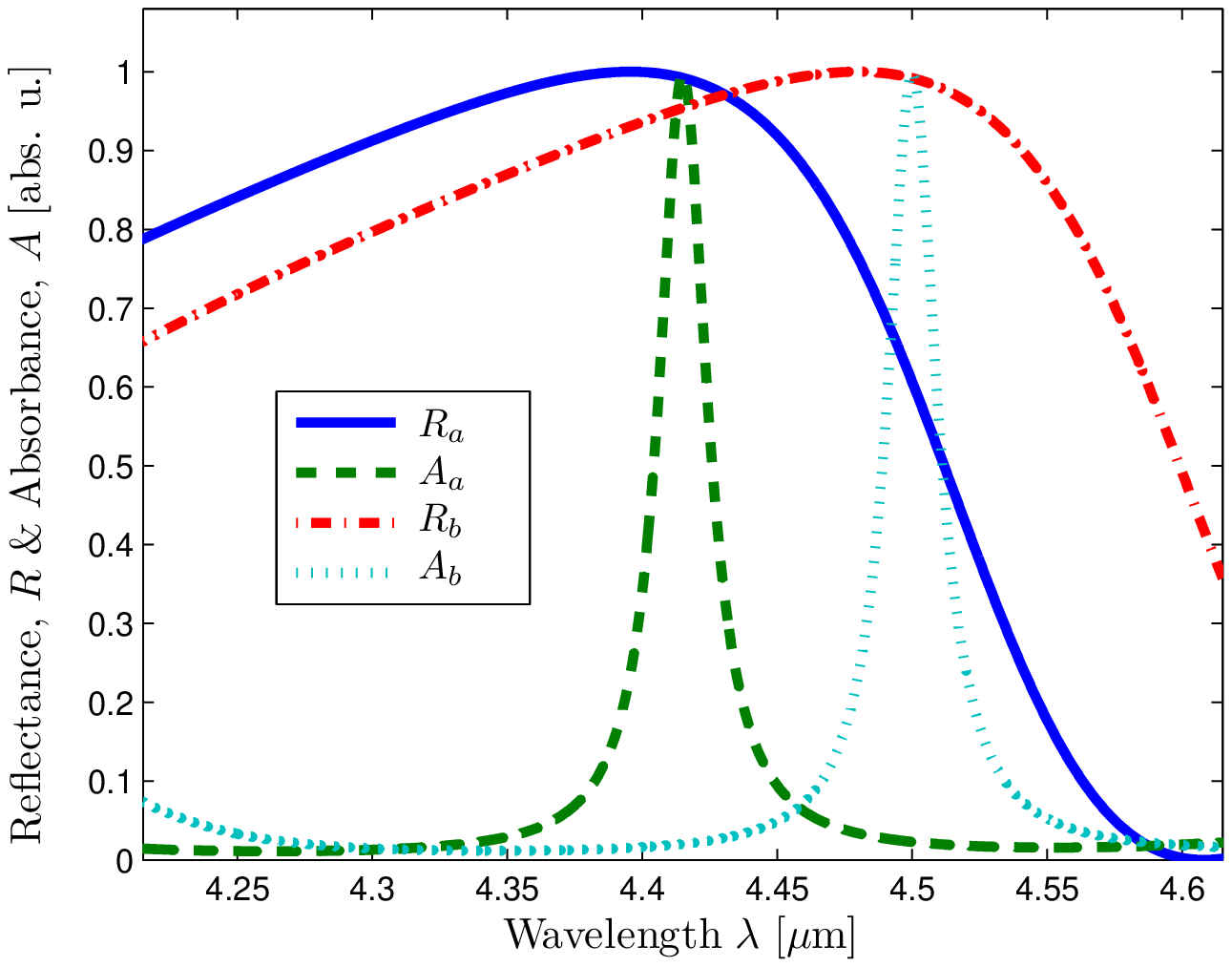}%0.38
      \caption{The spectra of reflectance from the front grating mirrors and of absorbance for the RCE-TE$_a$ and RCE-TE$_b$ structures.}
      \label{TEa&TEb RCE GMR and A}
    \end{center}
  \end{minipage}
  \hfill
\end{figure}
\end{center}
\vspace{-5pt}
Other trial dimensions were estimated semianalytically using the $\varphi_{\rm f}$ and $\varphi_{\rm b}$ as computed numerically for the stand-alone grating and DBR, respectively, and then refined by the optimization. The manual design was performed under the constraints: $d_{\rm f}=d_{\rm b}$, $w= W/ \Lambda = 0.5$ and $t_{\rm a}= 0.075\;\mu{\rm m}$, by controlling the $t_{\rm c}$ which allowed us to attain the desired $\delta_0$. To remind, the optimization goal is to achieve maximal polarized $A(\lambda_0)$, minimal changes of which due to variations in the grating fabrication process are being very desirable. A fully computer-aided optimization procedure easily allows for designing the structures with $d_{\rm f} \neq d_{\rm b}$ and $w\neq 0.5$. The results of such an optimization are the structures RCE-TE$_a$ and RCE-TM$_a$ ($\lambda_0 = 4.415\; \mu$m), RCE-TE$_b$ and RCE-TM$_b$ ($\lambda_0 = 4.500\; \mu$m) for operation with a TE and TM polarized radiation [see Table \ref{tab:t2}].

The TE and TM polarized $R$ spectra from the stand-alone grating mirrors and the related $A$ spectra of the RCE-TE$_a$ and RCE-TE$_b$ structures are shown in Fig. \ref{TEa&TEb RCE GMR and A} and \ref{TMa&TMb RCE GMR and A}, respectively. The designed gratings maintain the polarized $R \geq 99\%$ in a wide band around $\lambda_0 $. For the TM polarization, as seen in Fig. \ref{TMa&TMb RCE GMR and A}, the high-$R$ band width greatly exceeds $0.3\;\mu$m, in a good agreement with the paper \cite{chang-a}. For the TE polarization, however, this band is narrower which does not prevent from achieving the $A\approx 100$\% peak [see Fig. \ref{TEa&TEb RCE GMR and A}] and a superior tolerance to a variation in the grating fabrication process, as discussed in Subsect. \ref{subsect:RCE-toler} below.

%%---------------------------------------------------
\subsection{Design tolerances}
\label{subsect:RCE-toler}
An important issue for the practicality of the proposed RCE HgCdTe-PDs is the sensitivity to variations in the grating fabrication process. In this paper, we report the tolerance of the value and position of the $A$ peak to the grating groove duty cycle $w$ variations. The tolerance was found to be very good for the structures with the optimal $w\neq 0.5$, especially in the TE case. For example, the RCE-TE$_a$ structure still maintains high values of $A(\lambda_0) =93.8$\% and $98.8$\% with the initially prescribed $\lambda_0$ and $w=0.528$ and $w=0.728$, respectively. The small drop in $A(\lambda_0)$, should not be treated as a deterioration of the detection performance since the peak $A=99$\% appears to be shifted by at most 3 nm  from the $\lambda_0$, which indicates an excellent tolerance including a small change of the peak value and a minute shift of the peak position. The RCE-TE$_b$ structure exhibits similar tolerance properties.
\vspace{-12pt}
\begin{center}
\begin{figure}[h]
  \hfill
  \begin{minipage}[t]{.48\textwidth}
    \begin{center}
      \includegraphics[scale=0.38]{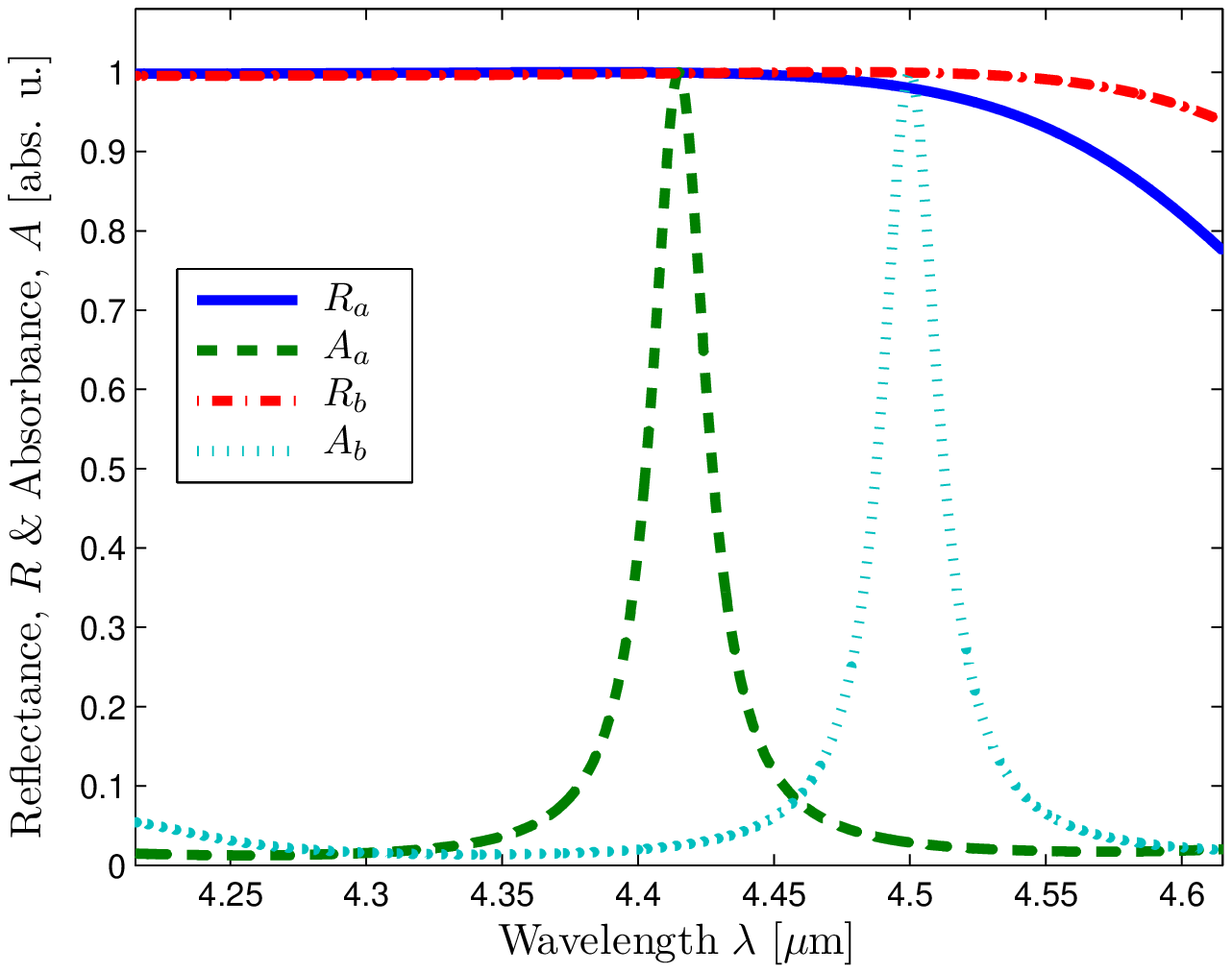}%0.38
      \caption{The reflectance from the stand-alone grating mirrors and the absorbance of the RCE-TM$_a$ and RCE-TM$_b$ structures.}
      \label{TMa&TMb RCE GMR and A}
    \end{center}
  \end{minipage}
  \hfill
  \begin{minipage}[t]{.48\textwidth}
    \begin{center}
      \includegraphics[scale=1.12]{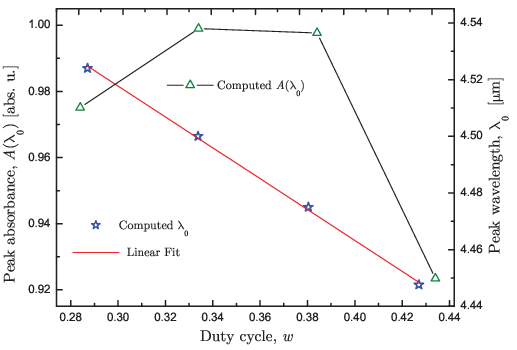}%1.14
      \caption{The peak absorbance and wavelength versus $w$ for the RCE-TM$_b$ structure.}
      \label{fig5}
    \end{center}
  \end{minipage}
  \hfill
\end{figure}
\end{center}
\vspace{-5pt}
The RCE-TM$_{a,b}$ structures withstand errors in $w$ of $\pm 0.05$ as related to peak $A$, which indicates a lower but still practical tolerance. However, the peak wavelength shift proves notably larger as compared to RCE-TE$_{a,b}$ structures. For example, as $w$ is shifted to $0.287$ the RCE-TM$_a$ structure exhibits a $24$ nm red shift of the absorption peak and a drop of the peak $A$ to $97.5\%$. In this case, we found numerically that the peak wavelength, within a reasonable range, depends on $w$ linearly, as shown in Fig. \ref{fig5}.

%%---------------------------------------------------
\section{Conclusion}
\label{sect:concl}
The RCE HgCdTe-PDs, including new ones in which the front mirror is a $\sim 100\%$ reflectance grating, for applications in MWIR range, were designed and simulated using semianalytic and computer-aided tools such as a scrutinized control of the resonant round-trip phase $\delta_0$, and the grating reflection amplitude and phase. The results show that the $\delta_0$ control is crucial for optimizing the efficiency, as was known for the conventional RCE PDs \cite{unlu,{faraone1},{faraone2},{faraone3}}. In our case, $n_{\rm a}$ and $n_{\rm c}$ differ significantly, but the reflections from the absorbing layer have a relatively small effect due to $t_{\rm a}\ll t_{\rm c}$. We also proved that a non-standard placement of the absorbing layer is capable of increasing the efficiency of the conventional RCE HgCdTe-PDs, which still remains below $85.2$\%.

It was decisively shown that for a linearly polarized light, the grating mirror based RCE HgCdTe-PDs can be designed to achieve $\sim 100\%$ efficiency, thus highly outperforming the conventional RCE HgCdTe-PDs, and simultaneously is highly tolerant to variations in the grating groove duty cycle $w$. For the TM designs, a linear dependence was found between the $w$ varied around its optimal value and the peak wavelength shifting, in a response, around the prescribed $\lambda_0$.
%-------------------------------------------------
\begin{table}[h]
\caption{The designed and simulated RCE HgCdTe-absorber structures. The dimensional parameters and $\lambda_{\rm{R}}$ are in microns. lowercase a and b refers to a wavelengths of $4.415 \;\mu$m and $4.500  \;\mu$m respectively.}
\label{tab:t2}
\begin{center}
\begin{tabular}{|c|c|c|c|c|c|c|c|c|c|}
%% |l|l| to left justify each column entry
%% |c|c| to center each column entry
%% use of \rule[]{}{} below opens up each row
\hline
\rule[-1ex]{0pt}{3.5ex}  Structures         & $d_{\rm{b}}$  & $t_{\rm{a}}$  & $d_{\rm{f}}$  & $t_{\rm{Ge}}$ & $t_{\rm{g}}$  & $\Lambda$ & $w$ & Peak $A$  & $\lambda_{\rm{R}}$\\%17.08.2011
\hline
\rule[-1ex]{0pt}{3.5ex}  RCE-S              & 1.177         & 0.075         & 1.177         & -             & -             & -         & -           & 0.819     & -\\%17.08.2011
\hline
\rule[-1ex]{0pt}{3.5ex}  RCE-O$_a$          & 0.272         & 0.075         & 0.433         & -             & -             & -         & -           & 0.831     & -\\%new 22.08.2011 4.415um
\hline
\rule[-1ex]{0pt}{3.5ex}  RCE-O$_b$          & 0.292         & 0.075         & 0.429         & -             & -             & -         & -           & 0.828     & -\\%new 22.08.2011 4.5um
\hline
\rule[-1ex]{0pt}{3.5ex}  RCE-TE$_a$         & 0.209         & 0.075         & 0.535         & 0.331         & 0.844         & 1.461     & 0.628       & 0.998     & 3.901\\%new 23.08.2011 4.415um
\hline
\rule[-1ex]{0pt}{3.5ex}  RCE-TE$_b$         & 0.228         & 0.075         & 0.534         & 0.335         & 0.861         & 1.492     & 0.629       & 0.999     & 3.984\\%new 23.08.2011 4.5um(maximum n_cavity=2.6703)
\hline
\rule[-1ex]{0pt}{3.5ex}  RCE-TM$_a$         & 0.259         & 0.075         & 0.504         & 0.628         & 0.921         & 1.456     & 0.337       & 0.999     & 3.888\\%new 23.08.2011 4.415um
\hline
\rule[-1ex]{0pt}{3.5ex}  RCE-TM$_b$         & 0.286         & 0.075         & 0.495         & 0.650         & 0.930         & 1.468     & 0.334       & 0.999     & 3.920\\%new 16.09.2011 4.5um(maximum n_cavity=2.6703)
\hline
\end{tabular}
\end{center}
\end{table}
%------------------------------------------------

\vspace{-26pt}


\begin{thebibliography}{}

%1
\bibitem{heavens} Heavens  O. S.: Optical Properties of Thin Solid Films, Dover Publ., New York (1991).

%2
\bibitem{knittl} Knittl Z.: Optics of Thin Films (an Optical Multilayer Theory, Wiley, New York (1976).

%3
\bibitem{unlu} \"{U}nl\"{u} M. S. and Strite S.: Resonant cavity enhanced photonic
devices, J. Appl. Phys., \textbf{78}, 607-639 (1995).

%4
\bibitem{deen} El-Batawy Y. M. and Deen M. J.: Resonant cavity enhanced
photodetectors (RCE-PDs): structure, material analysis and
optimization, Proc. SPIE, \textbf{4999}, 363-378 (2003).

%5
\bibitem{havaus} S. Hava and M. Auslender, Silicon grating-based mirror
for 1.3-$\mu$m polarized beams: MATLAB-aided design, Appl. Opt., \textbf{34}, 1053-1058 (1995).

%6
\bibitem{chang-a} C. F. R. Mateus, M. C. Y. Huang, Y. Deng,
A. R. Neureuther, and C. J. Chang-Hasnain, Ultrabroadband mirror using low-index
cladded subwavelength grating, IEEE Photonics Technol.
Lett., \textbf{16}, 518-520 (2004).

%7
\bibitem{bissaillon} E. Bissaillon, D. Tan, B. Faraji, A. G. Kirk, L. Chrostowski, and D. V. Plant,
``High reflectivity air-bridge subwavelength grating reflector and Fabry-Perot cavity in AlGaAs/GaAs,
Opt. Express, \textbf{14}, 2573-2582 (2006).

%8
\bibitem{kim}J. H. Kim, L. Chrostowski, E. Bisaillon, and D. V. Plant
DBR, subwavelength grating, and photonic crystal slab Fabry-Perot cavity design using phase analysis by FDTD, Opt. Express, \textbf{15}, 10330-10339 (2007).

%9
\bibitem{aushava1} M. Auslender and S. Hava, Scattering-matrix propagation algorithm in
full-vectorial optics of multilayer grating structures," Opt. Lett., \textbf{21}, 1765-1767 (1996).

%10
\bibitem{faraone1} J. G. A. Wehner, C. A. Musca, R. H. Sewell, J. M. Dell, and L. Faraone, Mercury
cadmium telluride resonant-cavity-enhanced photoconductive infrared detectors, Appl. Phys. Lett., \textbf{87}, 211104:1-3 (2005).

%11
\bibitem{faraone2} J. G. A. Wehner, C. A. Musca, R. H. Sewell, J. M. Dell, and L. Faraone, Responsivity and lifetime of
resonant-cavity-enhanced HgCdTe detectors, Solid-State Electron., \textbf{50}, 1640-1648 (2006).

%12
\bibitem{faraone3} J. G. A. Wehner, R. H. Sewell, J. Antoszewski, C. A. Musca, J. M. Dell, and
L. Faraone, Mercury cadmium telluride/cadmium telluride distributed Bragg reflectors for use with resonant cavity-enhanced
detectors, J. Electron. Mater., \textbf{34}, 710-715 (2005).

%13
\bibitem{kanev} J. Kaniewski, J. Muszalski and J. Piotrowski, Resonant
microcavity enhanced infrared photodetectors, Optica Applicata, \textbf{17}, 405-413 (2007).

%14
\bibitem{sioma} M. Sioma and J. Piotrowski, Modelling and optimisation of high
temperature detectors of long wavelength infrared radiation with
optically resonant cavity, Opto-Electron. Rev. \textbf{12}, 157-160 (2004).

\end{thebibliography}
\end{document}